\documentclass{article}
\usepackage{graphicx} 

\title{A note on the Moody diagram}
\author{Paulo R.\ de Souza Mendes\\
Department of Mechanical Engineering\\ 
Pontif\'icia Universidade Cat\'olica - RJ\\ 
Rio de Janeiro, RJ 22451-900 - Brazil\\
pmendes@puc-rio.br}

\begin{document}

\maketitle

\section{Introduction}

The steady flow of Newtonian fluids through tubes appears in a plethora of   applications, including various fields of engineering, physics and biology, to name a few. For this reason, it was the topic of several studies in the  nineteenth and  twentieth centuries. 

For laminar flow, G.\ H.\ L.\ Hagen (1797-1884) and 
J.\ L.\ M.\ Poiseuille (1797-1869) 
independently performed analytical and experimental studies that led to the same relation between the flow rate and the pressure drop, the so-called Hagen-Poiseuille equation.

For turbulent flow in tubes, prominent scientists like L. Prandtl, Th.\ von K\'arm\'an, J.\ Nikuradse, H.\ Darcy, H.\ Basin, C.\ Colebrook, and others performed experiments for wide ranges of Reynolds number and relative wall roughness.  Colebrook \cite{colebrook-39,colebrook-white-37} developed an equation for the friction factor (i.e.\ dimensionless pressure drop) as a function of the Reynolds number and relative roughness. 

Since this equation is transcendental, it was not very practical for routinely usage. 
In a celebrated work,  Moody \cite{moody-44} developed a diagram plotting both (a dimensionless version of) the Hagen-Poiseuille and the Colebrook equations. Moody's diagram  became a basic and indispensable tool that---since shortly after its publication nearly eighty years ago---has been routinely used by engineers and scientists worldwide in the design of a wide range of hydraulic systems. 

Other friction factor expressions for the turbulent flow in pipes have been developed over the years, some of them explicit
\cite{brkic-11,cojbasic-brkic-13,offor-alabi-16-2,offor-alabi-16,minhoni-pereira-silva-castro-saad-20,gallardo-rojas-guerra-21}.

In this work we propose an alternate scaling for the head loss  whose characteristics render more clear the role of inertia in this flow and ensure that the trends of the relationship between dimensionless quantities are the same ones observed in the dimensional problem. 

\section{Analysis}
A simple force balance in a fluid element in the steady, laminar, isochoric flow of a Newtonian fluid through a tube of constant cross sectional area ultimately leads to the well known Hagen-Poiseuille equation, that gives the volumetric flow rate $Q$ as a function of the pressure gradient $\Delta p/L$ ($\Delta p$ is the presssure difference between two axial positions separated by a distance $L$):
\begin{equation}
    Q=\frac{\pi D^4}{128\mu}\frac{\Delta p}{L}
    \label{eq:darcy-weisbach}
\end{equation}
where $D$ is the tube inner diameter and $\mu$ is the  viscosity of the fluid.

The most common dimensionless version of this equation is:
\begin{equation}
f=\frac{64}{Re} 
\label{eq:fre}
\end{equation}
where $f$ is the so-called friction factor, and $Re$ is the Reynolds number. These dimensionless quantities are defined as
\begin{equation}
    f:=\frac{(\Delta p/L)D}{\frac{1}{2}\rho \bar u^2}
    \label{eq:fdef}
\end{equation}
and 
\begin{equation}
    Re:=\frac{\rho\bar u D}{\mu}
    \label{eq:re}
\end{equation}
where $\rho$ is the fluid's density and $\bar u$ is the average axial velocity.

Since Eq.\ (\ref{eq:darcy-weisbach}) refers to a horizontal tube of constant cross sectional area, it is clear that $\Delta p$ is directly related to the mechanical energy loss that a Lagrangian particle of unit mass experiences as it travels along the  tube length $L$. Therefore, $f$ is the dimensionless head (or mechanical energy) loss.  

\begin{figure}
    \centering
\includegraphics[angle=0,width=0.9\textwidth]{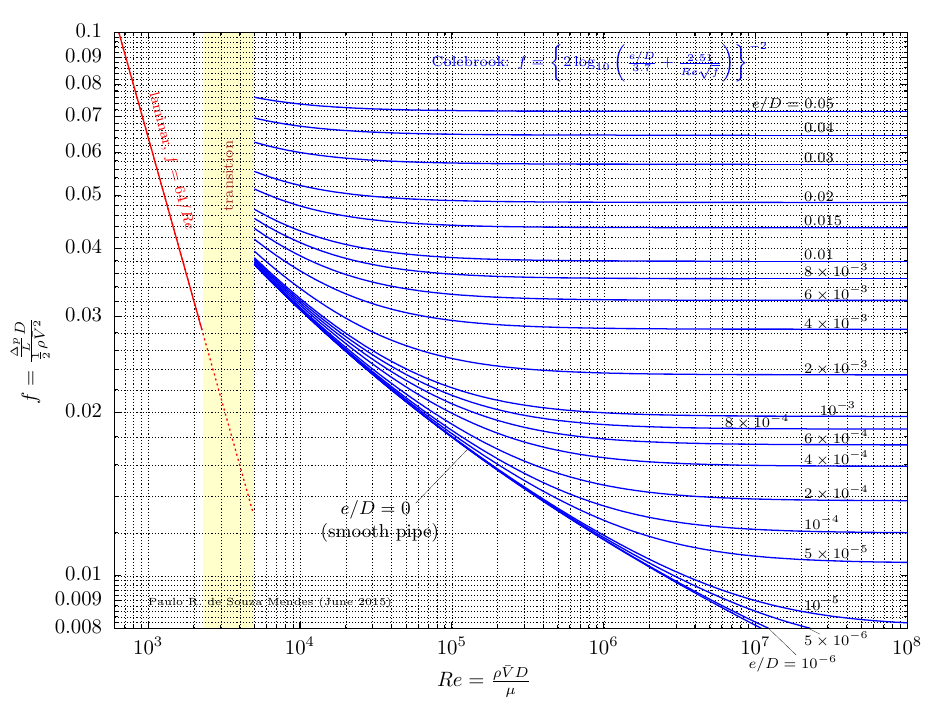}
    \caption{The classical Moody diagram}
    \label{fig:moody}
\end{figure}

For turbulent flows, the also famous Colebrook transcendental equation
\cite{colebrook-39} is employed, instead of Eq.\ (\ref{eq:fre}):
\begin{equation}
  f=\frac{1}{\left\{2\log_{10}\left(\frac{e/D}{3.7}+\frac{2.51}{Re\sqrt{f}}\right)\right\}^{2}} 
  \label{eq:colebrook}
\end{equation}
where $e$ is the average rugosity of the tube inner wall.
The transcendental nature of Eq.\ (\ref{eq:colebrook}) requires iterations to obtain $f$ as a function of $Re$ and $e/D$. However, the convergence is quite fast (2-4 iterations) and very weakly dependent of the initial guess for $f$, due to the appearance of $f$ on the RHS of Eq.\ (\ref{eq:colebrook}) within a square root which in turn is part of the argument of a logarithm. Nevertheless, its transcendental nature posed practical problems to last century users, who did not have access to efficient computer codes and spreadsheets. To address this problem, Moody \cite{moody-44} plotted Eqs.\  (\ref{eq:fre}) and (\ref{eq:colebrook}) in a diagram like the one shown in Fig.\ \ref{fig:moody}.

\section{Discussion}
A perhaps puzzling feature of the Moody diagram is the fact that the dimensionless  mechanical energy loss $f$ {\em decreases} as the dimensionless flow rate $Re$ is increased, both in the laminar and in the turbulent flow regimes.

The reason for this counterintuitive behavior is the fact that, in Eq.\ (\ref{eq:fdef}), an inertia force is used to non-dimensionalize the pressure force. However, in steady laminar flow in tubes, all material particles flow at constant velocity, so that there is no inertia force involved, the pressure force being exactly balanced by the viscous force. In turbulent flow, inertia plays an indirect role only, due to the velocity fluctuations around  average constant values experienced by the material particles. 

Therefore, it seems more appropriate to use, in the non-dimensionalization of the pressure force, a characteristic {\em viscous  force} rather than an inertia force. The result is a modified friction factor $f^*$, defined as
\begin{equation}
f^*:=\frac{D^2}{32\mu\,\bar u}\frac{\Delta p}{L}
    \label{eq:fstardef}
\end{equation}
Combining Eqs.\ (\ref{eq:darcy-weisbach}) and (\ref{eq:fstardef}), we get, for laminar flow,

\begin{equation}
   f^*=1 
\label{eq:fstar}
\end{equation}

Regarding turbulent flow, is not difficult to re-write the Colebrook equation (Eq.\ (\ref{eq:colebrook})) in terms of the modified friction factor:

\begin{equation}
   f^*= \frac{Re}{\left\{16\log_{10}\left(\frac{e/D}{3.7}+\frac{0.314}{\sqrt{Ref^*}}\right)\right\}^{2}} 
  \label{eq:colebrookstar}
\end{equation}

\begin{figure}
    \centering
\includegraphics[angle=0,width=0.9\textwidth]{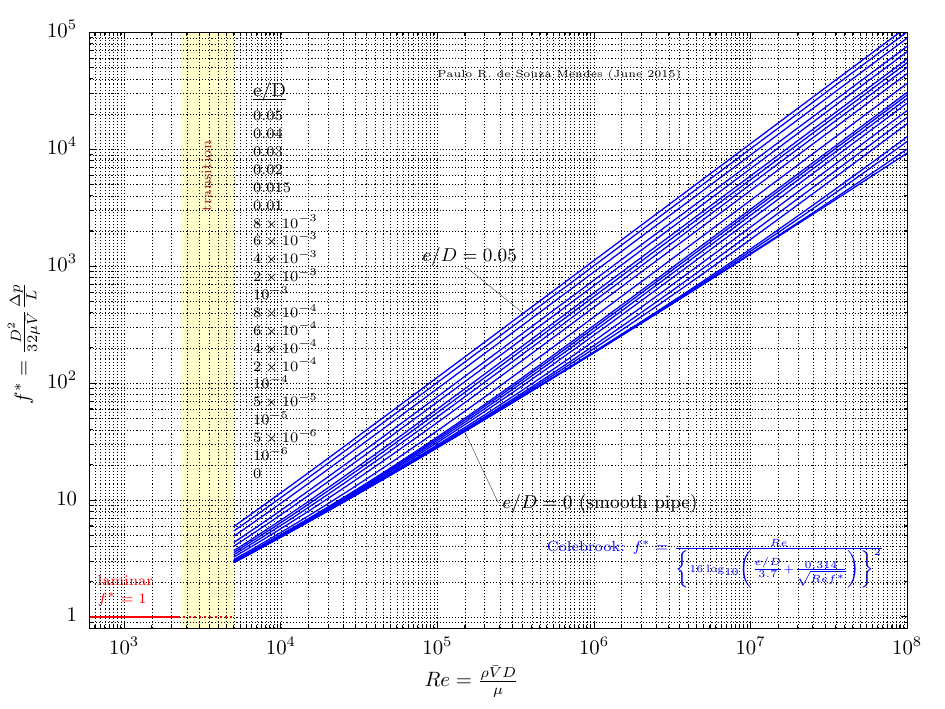}
    \caption{The modified Moody diagram}
    \label{fig:moodyfied}
\end{figure}
We now represent Eqs.\ (\ref{eq:fstar}) and (\ref{eq:colebrookstar}) in a modified version of the Moody diagram, shown in Fig.\ \ref{fig:moodyfied}. In this figure we observe that the friction factor is constant for laminar flow, and increases monotonically with the Reynolds number for turbulent flow. This behavior observed for the non-dimensional quantities is in agreement with the one expected for the dimensional quantities, because the new scaling is more consistent with the physics involved in this problem.

\section{Concluding remarks}
In this brief note we exemplify the importance of choosing the appropriate scaling to obtain dimensionless quantities that are faithful to their dimensional counterparts, thus aiding the understanding of the physics involved.

Of course this alternate non-dimensionalization has no advantage over the traditional one as far as the solution of practical problems is concerned.


\bibliographystyle{plain}
\end{document}